# Direct Imaging of Sketched Conductive Nanostructures at the LaAlO$_3$/SrTiO$_3$ Interface

Zhanzhi Jiang[1], Xiaoyu Wu[1], Hyungwoo Lee[3], Jung-Woo Lee[3], Jianan Li[2,4], Guanglei Cheng[2,4,5], Chang-Beom Eom[3], Jeremy Levy[2,4], Keji Lai[1]

[1]Department of Physics, University of Texas at Austin, Austin TX 78712, USA

[2]Department of Physics and Astronomy, University of Pittsburgh, Pittsburgh PA 15260, USA

[3]Department of Materials Science and Engineering, University of Wisconsin-Madison, Madison, WI 53706, USA

[4]Pittsburgh Quantum Institute, Pittsburgh, PA 15260, USA

[5]CAS Key Laboratory of Microscale Magnetic Resonance and Department of Modern Physics, University of Science and Technology of China, Hefei 230026, China

## Abstract

Nanoscale control of the quasi-two-dimensional electron gas at the LaAlO$_3$/SrTiO$_3$ (LAO/STO) interface by a conductive probe tip has triggered the development of a number of electronic devices. While the spatial distribution of the conductance is crucial for such devices, it is challenging to directly visualize the local electrical properties at the buried interface. Here we demonstrate conductivity imaging of sketched nanostructures at the LAO/STO interface by microwave impedance microscopy (MIM) with a lateral resolution on the order of 100 nm. The sheet conductance extracted from the MIM data agrees with the transport measurement. The tip-induced insulator-to-metal transition is observed above a threshold voltage of +4 V. Our work paves the way to study emergent phenomena at oxide interfaces by probing nanoscale conductance distribution.

The interface between two insulating perovskites LaAlO$_3$ (LAO) and SrTiO$_3$ (STO) has been in the limelight of material research in the past decade [1]. When 4 or more unit cells (uc) of LAO are epitaxially grown on a TiO$_2$-terminated STO substrate, a high-mobility quasi-two-dimensional electron gas (q2DEG) forms spontaneously on the STO side of the interface [2]. This system exhibits a plethora of intriguing phenomena including the presence and coexistence of superconductivity and ferromagnetism [3-6]. At the critical LAO thickness of 3 uc, a metastable metal-insulator transition can be controlled using either a global gate voltage applied on the back of the STO substrate [7] or on the top LAO surface using a conductive atomic-force microscopy (c-AFM) tip [8]. In the latter case, the q2DEG can be reversibly written and erased underneath the tip with nanoscale lateral dimensions, enabling the creation and control of a variety of nanostructures such as sketch field effect transistors (FETs) [8-10], photodetectors [11], single electron transistors [12], and quantum dots [13, 14].

Scanning probe microscopy (SPM) has played a key role in understanding the rich physics at the LAO/STO interface. Aside from the aforementioned c-AFM nano-patterning [8-14], the depth profile of the q2DEG has been visualized by cross-sectional c-AFM imaging [15], and the surface charge distribution and electromechanical response have been imaged by electric force microscopy (EFM) [16] and piezo-force microscopy (PFM) [17, 18], respectively. Other properties of the LAO/STO system, including the electrostatic potential, magnetism, and superconductivity, are also locally probed using scanning single-electron transistor (SET) microscopy [19] and scanning superconducting quantum interference device (SQUID) microscopy [6, 20]. On the other hand, one of the most important physical quantities in this system, the local 2D conductivity, has not been directly imaged in a quantitative manner. In this work, we report the patterning and imaging of sketched conductive nanostructures at the 3uc-LAO/STO interface by microwave impedance microscopy (MIM) [21]. The MIM is a powerful tool for sub-surface conductivity imaging with a lateral resolution on the order of 100 nm [22]. The sketched q2DEG patterns can be visualized by the MIM, and the estimated sheet resistance can be extracted using finite-element analysis [23], in agreement with the transport data. The local insulator-to-metal transition is also demonstrated as a function of the tip voltage. Our work lays the foundation to explore various emergent phenomena in oxide interfaces with local electrical probes.

The LAO/STO sample in this experiment is prepared as follows. The LAO thin film is deposited on a TiO$_2$-terminated STO (001) substrate by pulsed laser deposition with *in situ* high-pressure reflection high-energy electron diffraction (RHEED) monitoring. The 3 uc of LAO film is grown at a temperature of 550 °C and O$_2$ pressure of $1\times10^{-3}$ mbar [24]. After growth, electrical contacts to the interface are prepared by Ar-ion milling 25 nm deep trenches and filling them with Au/Ti bilayer (2 nm adhesion Ti layer and 23 nm Au layer).

The MIM measurement is performed on an AFM platform (XE-70, Park Systems) [21]. As shown in Fig. 1a, an excitation signal (10 µW at 1 GHz) is fed to the center conductor of a shielded cantilever tip [22] and the reflected microwave is demodulated to produce the two output signals MIM-Im and MIM-Re, which are proportional to the imaginary and real parts of the tip-sample admittance, respectively. The GHz frequency is essential for effective capacitive coupling to the buried q2DEG at the LAO/STO interface, as well as to suppress the piezoelectric modes as probed in PFM [17, 18]. The equivalent lump-element circuit of the tip-sample interaction is sketched in the inset. The impedance between the tip and the ground is dominated by the STO substrate when the LAO/STO interface is insulating, and by the LAO layer when the interface is highly conductive interface. In addition to the microwave signals, a DC bias $V_{tip}$ is also applied to the tip through a bias-tee. The two-terminal conductance between the Au/Ti electrodes is monitored by a Keithley source meter with $V_{DS}$ = 10 mV. As an extension of the metal contacts, two rectangular pads are first written with $V_{tip}$ = +5 V. A nanowire is then drawn by scanning the MIM tip with the same bias from one pad to the other. Results shown here are representative of over 20 nanowires created in this fashion using several MIM tips. The abrupt increase of conductance due to the connection of two pads is around 0.1 ~ 1 µS for each micrometer in length of the wire. Similar to previous reports [24], the conductance slowly decreases under the ambient condition, with a characteristic time scale (several hours) much longer than the MIM imaging time (~ 10 min). We note that this GHz imaging technique cannot distinguish conductivity arising from "water cycle" [24] versus oxygen vacancy formation [25]. The AFM and MIM images of a typical nanowire are shown in Figs. 1b and 1c, respectively. The step-terrace-like features in the MIM-Im images are due to the topographic crosstalk [22]. On the other hand, with no corresponding observable features in the surface topography, the sketched wire is clearly seen in the MIM data. Note that the two-terminal conductance remains stable during the imaging, indicative of the noninvasive nature of the MIM measurement. The nanowire is then cut by moving the tip with $V_{tip} = -5$ V perpendicular to the wire, accompanied by a sudden drop of the current [8]. The MIM images (Fig. 1d) taken after the erasing process also show a clear breakage at the cutting point.

The MIM line profiles of the nanowire in Fig. 1c are plotted in Fig. 2a. For the > 20 wires measured in this study, the full-width-half-maximum (FWHM) linewidths are around 150 ~ 250 nm, which is limited by the spatial resolution [21] rather than the actual width of the nanowire. We note that the MIM signal levels, defined as the contrast between the peak values and the insulating background, are different for different tips in the measurements. On the other hand, it has been shown that the ratio between MIM-Im and MIM-Re signals is much less affected by the uncertainty on tip shapes [26]. In the following, we will compare the MIM-Im/Re ratio with the modeling results. To estimate the local conductivity of the nanowire, we simulate the tip-sample admittance by finite-element analysis (FEA) [23] using commercial software COMSOL 4.4. Here the 3 uc LAO is modeled as a 1.2-nm-thick layer with a relative permittivity of 25 [27]. The relative permittivity of the STO substrate is 300. The nanowire located on the STO side of the interface is assumed to be 10 nm in width [8]. The radius of the MIM tip in the simulation is 100 nm, consistent with the scanning electron microscopy (SEM) image of a typical probe (inset of Fig. 2a) and the FWHM in the MIM line profiles. Under these conditions, the simulated MIM signals [23] as a function of the sheet resistance $R_{sh}$ of the nanowire are plotted in Fig. 2b, from which $R_{sh}$ ~ 30 kΩ/sq can be estimated by comparing the measured MIM-Im/Re ratio (around 2:1 for the nanowires) with the COMSOL results.

In the single-wire experiment described above, it is difficult to directly compare the MIM and transport results because of the uncertainty in the nanowire width and the contact resistance. To circumvent these problems, we write multiple wide bands, each about 15 squares, with $V_{tip} = +5$ V across two pre-patterned rectangular pads extended from a pair of electrodes. It is worth noting that the lateral dimension of the conductive region is now greater than the tip diameter. The sketched conductive features are clearly observed in Fig. 3a, with measured MIM-Im/Re ratio around 4:1, as seen in the line profiles in Fig. 3b. The conductance after the writing of each band is recorded. Assuming that the contact resistance $R_{contact}$ and the background leakage resistance $R_{bg}$ of the device stay constant throughout the measurement, we can express the total resistance across the two electrodes as follows.

$$R_{\text{total}} = 2R_{\text{contact}} + \left(R_{\text{bg}}^{-1} + n \cdot R_{\text{band}}^{-1}\right)^{-1} \quad (1)$$

As shown in Fig. 3c, the experimental data fit nicely to Eq. (1) with a resistance of each band $R_{\text{band}} \sim 300$ k$\Omega$ and the corresponding sheet resistance $R_{\text{sh}} \sim 20$ k$\Omega$/sq. Fig. 3d shows the simulated MIM signals when the size of the conductive region is much larger than the tip diameter. By comparing the experimental data and FEA results, the sheet resistance extracted from the MIM images is 20 ~ 30 k$\Omega$/sq, which is in excellent agreement with the transport data.

Finally, we demonstrate the visualization of the insulator-to-metal transition at the LAO/STO interface by MIM imaging. Fig. 4a shows the MIM signals during repeated line scans across two pads when $V_{\text{tip}}$ ramps from 0 to +5 V. As seen from the selected line profiles in Fig. 4b, the MIM signals on the pads remain constant during this process, whereas the signals on the nanowire region rise as increasing $V_{\text{tip}}$. As illustrated in the insets of Fig. 2b and Fig. 3d, the conductive region underneath the tip is much narrower in the case of a nanowire than that of a pad. As a result, for the same sheet conductance in these two regions, the MIM-Im signals (proportional to the tip-sample capacitance) will be larger on the pad than that on the wire. Moreover, we note that the MIM-Re peaks at the nanowire-pad junctions are not fully understood. Using the same FEA simulation described above, it is straightforward to show that the conductance across the two pads emerges rapidly once the tip bias exceeds a threshold voltage $V_{\text{th}} \sim +4$ V, as plotted in Fig. 4c. This threshold voltage is consistent with previous c-AFM studies [9], as well as the observation that surface charge accumulation on the LAO surface is enhanced for $V_{\text{tip}} > +4$ V [16]. The result suggests that the interfacial conductance is due to the field effect through charge writing at the sample surface. We emphasize that the measurement is made possible by the ability of conductivity imaging in the presence of a tip bias.

In conclusion, using a microwave impedance microscope, we have demonstrated the non-invasive visualization of conductive nanostructures at the 3uc-LAO/STO interface. The MIM not only reveals the sketched patterns but also provides a quantitative measurement on the local sheet resistance of the q2DEG, which is in good agreement with the transport data. The insulator-to-metal transition is observed at a threshold tip bias of +4 V, beyond which the local conductivity is strongly enhanced. Our results provide opportunities to study the emergent phenomena at various oxide interfaces [28, 29].

## Acknowledgement

The MIM experiment at University of Texas at Austin was supported by the US Department of Energy (DOE), Office of Science, Basic Energy Sciences, under the Award DE-SC0010308. Data analysis was sponsored by the Army Research Office and was accomplished under Grant Number W911NF-17-1-0542. J.L. acknowledges support from the Vannevar Bush Faculty Fellowship program sponsored by the Basic Research Office of the Assistant Secretary of Defense for Research and Engineering and funded by the Office of Naval Research through grant N00014-15-1-2847. The work at University of Wisconsin-Madison was supported by AFOSR through grant FA9550-15-1-0334, AOARD FA2386-15-1-4046 and the National Science Foundation under DMREF Grant No. DMR-1629270. The views and conclusions contained in this document are those of the authors and should not be interpreted as representing the official policies, either expressed or implied, of the Army Research Office or the U.S. Government. The U.S. Government is authorized to reproduce and distribute reprints for Government purposes notwithstanding any copyright notation herein.

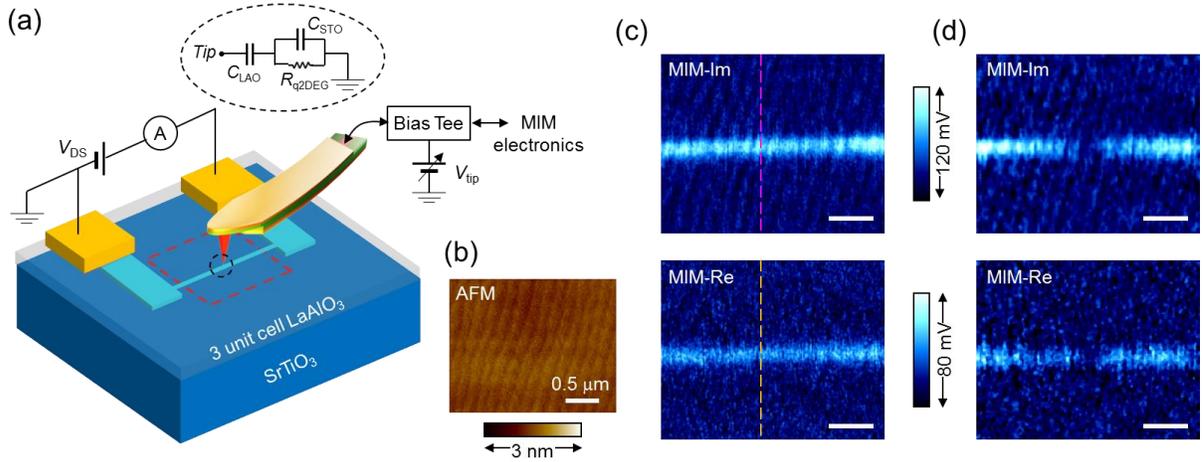

FIG. 1. (a) Schematic of the experimental setup. Both the microwave excitation and the tip bias are applied to the shielded cantilever tip through a bias-tee. The two-terminal conductance is monitored by a source-drain bias across the Ti/Au electrodes. The inset shows the equivalent lump-element circuit of the tip-sample interaction. Here $C_{LAO}$, $C_{STO}$, and $R_{q2DEG}$ represent the capacitance of the LAO layer, STO substrate, and resistance of the q2DEG layer, respectively. (b) AFM image inside the dashed rectangle in (a). (c) MIM images in the same area as (b) after the writing of a nanowire with $V_{tip} = +5$ V. (d) MIM images after the wire is cut in the middle by a tip bias of -5 V. All scale bars in (b – d) are 0.5 μm.

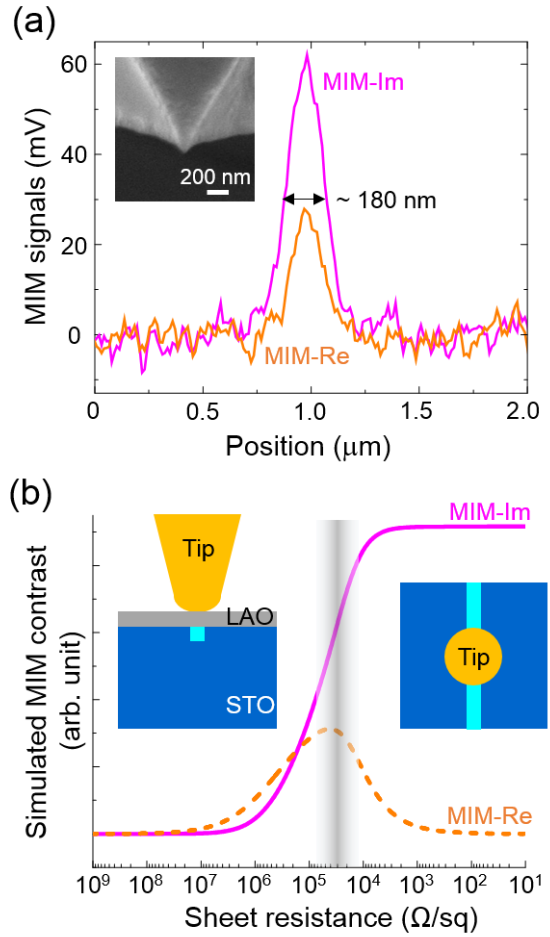

FIG. 2. (a) MIM line profiles of the nanowire labeled as the dashed lines in Fig. 1c. The inset shows the SEM image of a typical MIM tip. (b) FEA simulation of the MIM contrast (with respect to the insulating background) as a function of the sheet resistance for a 10 nm wire underneath the tip. The signal levels from (a) are consistent with $R_{sh} \sim 30$ k$\Omega$/sq, as indicated in the shaded area. The insets show the side (left) and top (right) views of the tip-sample configuration in the simulation.

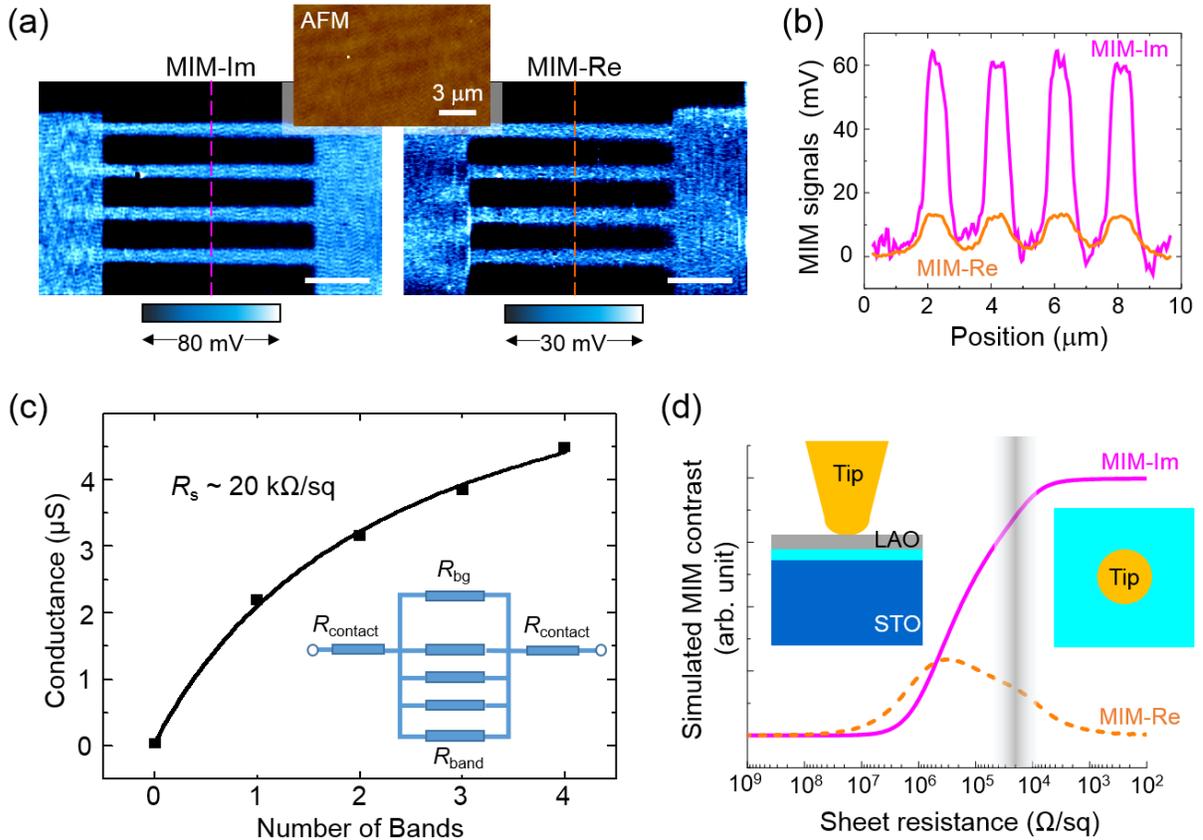

FIG. 3. (a) AFM and MIM images of four bands across two pads, all written with $V_{tip} = +5$ V. The scale bars are 3 μm. (b) MIM line profiles across the bands labeled as the dashed lines in (a). (c) Conductance between the electrodes as a function of the number of bands. The solid line is a fit to the experimental data (black squares) using a sheet resistance $R_{sh} \sim 20$ kΩ/sq. (d) Simulated MIM signals as a function of $R_{sh}$ when the conductive region is much wider than the tip diameter. The insets show the side (left) and top (right) views of the tip-sample configuration in the simulation. The contrast in (b) is consistent with $R_{sh} \sim 20$ kΩ/sq, in good agreement with the transport data.

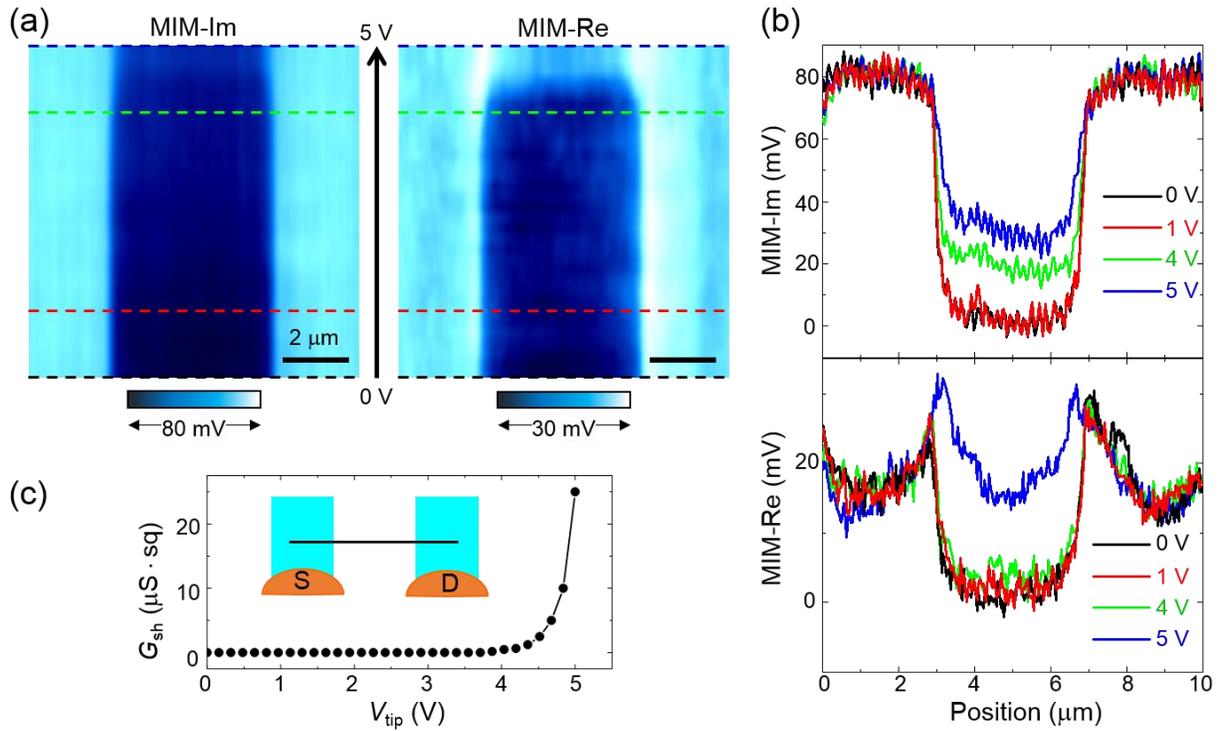

FIG. 4. (a) MIM signals as the tip repeatedly scans between two pads. The y-axis in the plots represent $V_{tip}$, which ramps from 0 (bottom) to +5 V (top). The scale bars are 2 μm. (b) Selected MIM line profiles labeled as the dashed lines in (a). (c) Sheet conductance $G_{sh}$ as a function of $V_{tip}$ estimated from the FEA. The conduction at the LAO/STO interface becomes evident above a threshold voltage of +4 V. The inset shows a schematic of the repeated line scans (black line).

12